\documentclass{iaus}
\usepackage{graphicx}

\title[An A star on an M star during a flare within a flare] 
{An A star on an M star during a flare within a flare}

\author[Kowalski, A. F. et al]   
{Adam F. Kowalski,$^1$
  Suzanne L. Hawley,$^1$ Jon A. Holtzman,$^2$ John P. Wisniewski,$^{1,3}$ Eric J. Hilton$^1$}

\affiliation{$^1$Astronomy Department, University of Washington \\ Box 351580, Seattle, WA 98195, USA
 \\ email: {\tt kowalski@astro.washington.edu} \\[\affilskip]
$^2$Department of Astronomy, New Mexico State University\\ Box
30001, Las Cruces, NM 88003, USA \\[\affilskip]
$^3$NSF Astronomy \& Astrophysics Postdoctoral Fellow}

\pubyear{2011}
\volume{273}  
\pagerange{1--4}
\jname{Physics of Sun and Star Spots}
\editors{D.P. Choudhary, \& K.G. Strassmeier, eds.}
\begin{document}

\maketitle

\begin{abstract}
M dwarfs produce explosive flare emission in the near-UV and optical continuum, 
and the mechanism responsible for this phenomenon is not well-understood.  We
present a near-UV/optical flare spectrum from the rise phase of a secondary flare, which occurred 
during the decay of a much larger flare.  The newly formed flare emission resembles the spectrum
of an early-type star, with the Balmer lines and continuum in \emph{absorption}.  
We model this observation phenomonologically as a temperature bump (hot spot) near 
the photosphere of the M dwarf.  The amount of heating implied by our model  ($\Delta T_{phot} \sim 16,000$ K) 
is far more than predicted by chromospheric backwarming in current 1D RHD flare models ($\Delta T_{phot} \sim 1200$ K).  

\end{abstract}

\firstsection 
\section{Introduction}
Flares on M dwarfs are notorious for producing dramatic outbursts in the near-UV and 
optical (white light) continuum.
The white light continuum has been observed in both the impulsive and
decay phases of stellar flares, and the
broadband shape of this emission resembles that of a hot blackbody with $T\sim8500-10,000$ K
(\cite[Hawley \& Fisher 1992]{Hawley1992}, \cite[Hawley et al. 2003]{Hawley2003}). In contrast, radiative hydrodynamic (RHD) flare
models predict a white light continuum with a prominent Balmer continuum in emission (\cite[Allred et al. 2006]{Allred2006}).  However,
when convolved with broadband filters the model spectrum \emph{also} exhibits the shape of a hot 
blackbody (\cite[Allred et al. 2006]{Allred2006}).
Spectra have been obtained during M dwarf flares (\cite[Hawley \& Pettersen 1991]{Hawley1991}, \cite[Eason et al. 1992]{Eason1992}, \cite[Garcia-Alvarez et al. 2002]{GarciaAlvarez2002}, \cite[Fuhrmeister et al. 2008]{Fuhrmeister2008}), showing a clear rise
into the near-UV without an abrupt discontinuity at the Balmer jump wavelength ($\lambda=3646$ \AA) 
or a prominent Balmer continuum in emission.  To unravel
this complexity in the white light continuum, we have begun a detailed investigation at wavelengths 
near the Balmer jump using new, time-resolved flare spectra and models.


On UT 2009 January 16, we observed an incredible flare on the dM4.5e star YZ CMi, obtaining high-cadence near-UV/optical (3350$-$9260 \AA) spectra and simultaneous $U$-band photometry from the ARC 3.5-m and NMSU 1-m telescopes at APO 
(see \cite[Kowalski et al. 2010]{Kowalski2010}; hereafter K10).  The spectra cover the section of the flare decay shown in Figure 1(a), 
which consists of several secondary flare peaks. 
The K10 analysis
revealed two continuum components in the flare spectra:  a Balmer continuum in emission as predicted by 
the RHD models
and a hot ($T$$\sim$10,000 K), compact blackbody as seen in previous flare observations.

\clearpage
\begin{figure}[b]
\begin{center}
 \includegraphics[width=2.8in,angle=90]{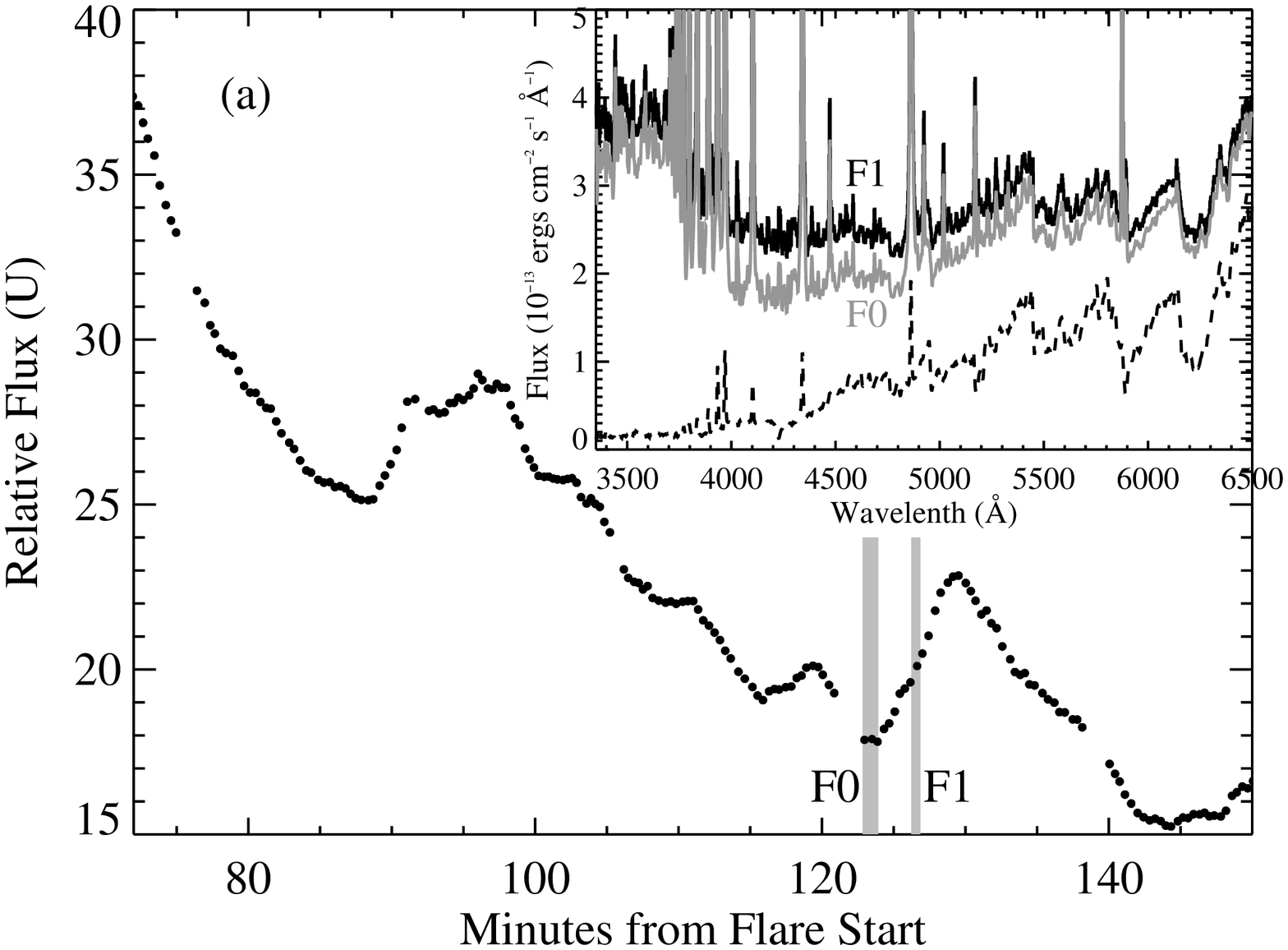} 
\includegraphics[width=2.8in,angle=90]{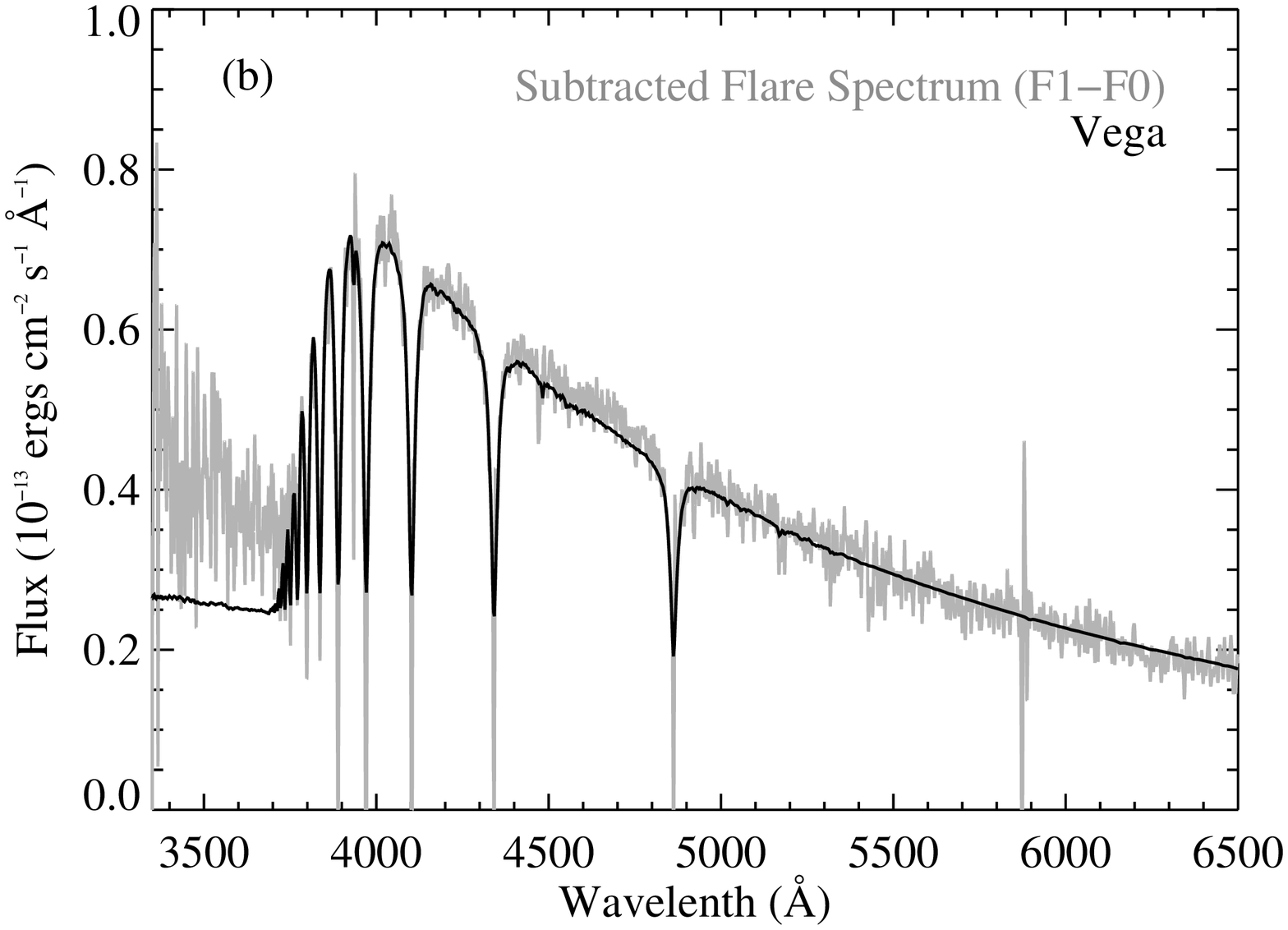} 
 \caption{(a) The $U$-band photometry is shown for the time during which spectra were obtained (see
K10 for the complete light curve of the flare).  The inset displays averaged spectra immediately before a secondary
flare (grey, F0), 2.2 min into the rise phase (black, F1), and during quiescence (dashed line).   The times 
encompassed by each of the flare spectra
 are denoted by vertical grey 
bars in the $U$-band light curve.  (b) Subtracting the F0 spectrum from the F1 spectrum isolates newly formed flare emission (grey) during the secondary flare rise.  The resulting subtracted flare spectrum resembles the A star Vega with the Balmer features in absorption.  }
   \label{fig1}
\end{center}
\end{figure}

\vspace*{-0.6 cm}
\section{The Secondary Flare Spectra}
During the secondary flares, K10 found that the time evolution of the Balmer 
continuum and Balmer line fluxes are anti-correlated with the areal coverage of
the $T\sim10,000$ K blackbody component.  For example, a secondary flare rise begins
at $t \sim 124$ min and lasts $\sim$5 min, resulting in an apparent decrease by 40\% 
in the Balmer continuum flux and an increase by a factor of nearly 2 in 
the area of the blackbody-emitting region. 


The total flare spectra at two times
are shown in the inset in Figure 1(a).  The times correspond
to immediately before the secondary flare rise (F0; $t = 123.4$ min) and 2.2 min
into the 
secondary flare rise (F1; $t = 126.5$ min). The
 spectra are clearly very complex, consisting of line and continuum emission from both 
 previously 
heated and newly formed flare regions, in addition to molecular
band absorption from the surrounding non-flaring photosphere.  We isolate the newly formed flare emission by 
subtracting the F0 spectrum from the F1 spectrum.
The striking features of this subtracted flare spectrum (Figure 1b)
 include a Balmer continuum and lines in absorption\footnote{We use the term \emph{absorption} throughout to refer to  `less emission than the neighboring spectral regions'.}, in contrast
to the total flare spectrum in which the Balmer continuum is in emission. Using continuum windows from 4000$-$6500 \AA,  the spectrum is fit by a blackbody with $T \sim$ 16,500 K.
We show the spectrum of the A0 V star Vega (from \cite[Bohlin 2007]{Bohlin2007}) 
in Figure 1(b) to highlight the remarkably similar characteristics with the spectrum of an early-type star.  

\vspace*{-0.6 cm}
\section{Phenomonological Modelling with RH}
We model the
 flare spectrum by placing a hot spot,
represented by a Gaussian temperature bump with peak temperature 
of 20,000 K, deep (log$_{10}$ col mass $ = 0.5$ g cm$^{-2}$) in the quiescent M dwarf 
atmosphere (Figure 2a).  The emergent radiation is calculated with the static NLTE code RH (\cite[Uitenbroek 2001]{Uitenbroek2001}) with a 5 level (plus continuum) Hydrogen atom.  We initially iterate to 
solve for the NLTE background opacities for Hydrogen, and we also consider some relevant molecular species 
that include Hydrogen (e.g., H$_2$).
The emergent hot spot spectrum is shown in Figure 2(b).  The model
spectrum has a
blackbody temperature of $\sim$18,000 K, and the Balmer continuum and lines are 
in absorption, similar to the subtracted flare spectrum.  In future work, we will
include Helium and metallic transitions, additional levels in the Hydrogen atom, and refined electron densities.  We will
also further constrain the parameters and time evolution of the hot spot.

\vspace*{-0.6 cm}
\section{Summary \& Discussion}
We find that newly formed emission during a secondary flare resembles the spectrum
of an early-type star, such as Vega.  Modelling the spectrum phenomonologically\footnote{We note that our
phenomonological flare models are very similar 
to the semi-empirical models for the absorption profiles seen during during Ellerman Bombs
on the Sun (e.g., \cite[Fang et al. 2006]{Fang2006}).}
by placing a hot spot near the quiescent M dwarf photosphere adequately reproduces 
the observed spectral shape and Hydrogen absorption features.  In a future work, we will show
how combining all flare continuum components reproduces the \emph{total} flare
spectrum, thereby providing an explanation for the anti-correlation in the time
evolution between the Hydrogen emission
and blackbody components.

Current, self-consistent 1D RHD models 
of stellar flares predict that the photosphere is 
heated by only $\sim$1200 K, predominantly due to backwarming from the flare chromosphere 
(\cite[Allred et al. 2006]{Allred2006}). 
Heating of deep layers by the amount needed to produce a $T_{max}$$\sim20,000$ K hot spot is clearly
not achieved by a solar-type non-thermal electron beam as energetic as 10$^{11}$ ergs cm$^{-2}$ s$^{-1}$.  
More spectral 
observations of flares with time resolution much less than the rise time, and with spectral coverage 
farther into the near-UV, are needed to characterize the
ubiquity of the phenomena presented in this work.

 \vspace*{-0.6 cm}
\section{Acknowledgements}
The presentation of this paper in the IAU Symposium 273 was possible due to partial 
support from the National Science Foundation grant numbers ATM 0548260, AST 0968672, 
AST 0807205 and NASA - Living With a Star grant number 09-LWSTRT09-0039.  We thank
H. Uitenbroek for his helpful discussions about RH.



\begin{figure}[b]

 \includegraphics[width=1.9in,angle=90]{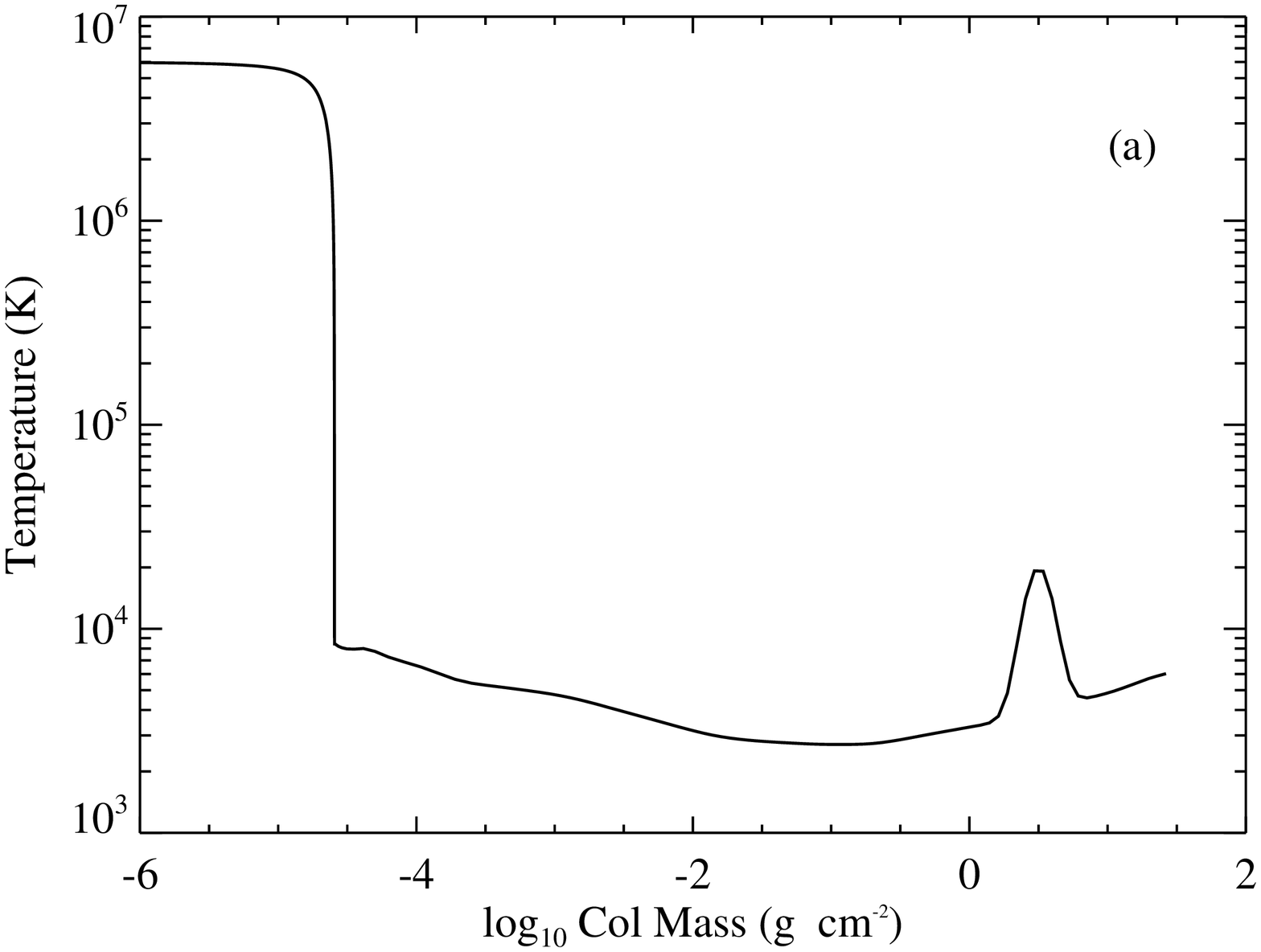}
\includegraphics[width=1.9in,angle=90]{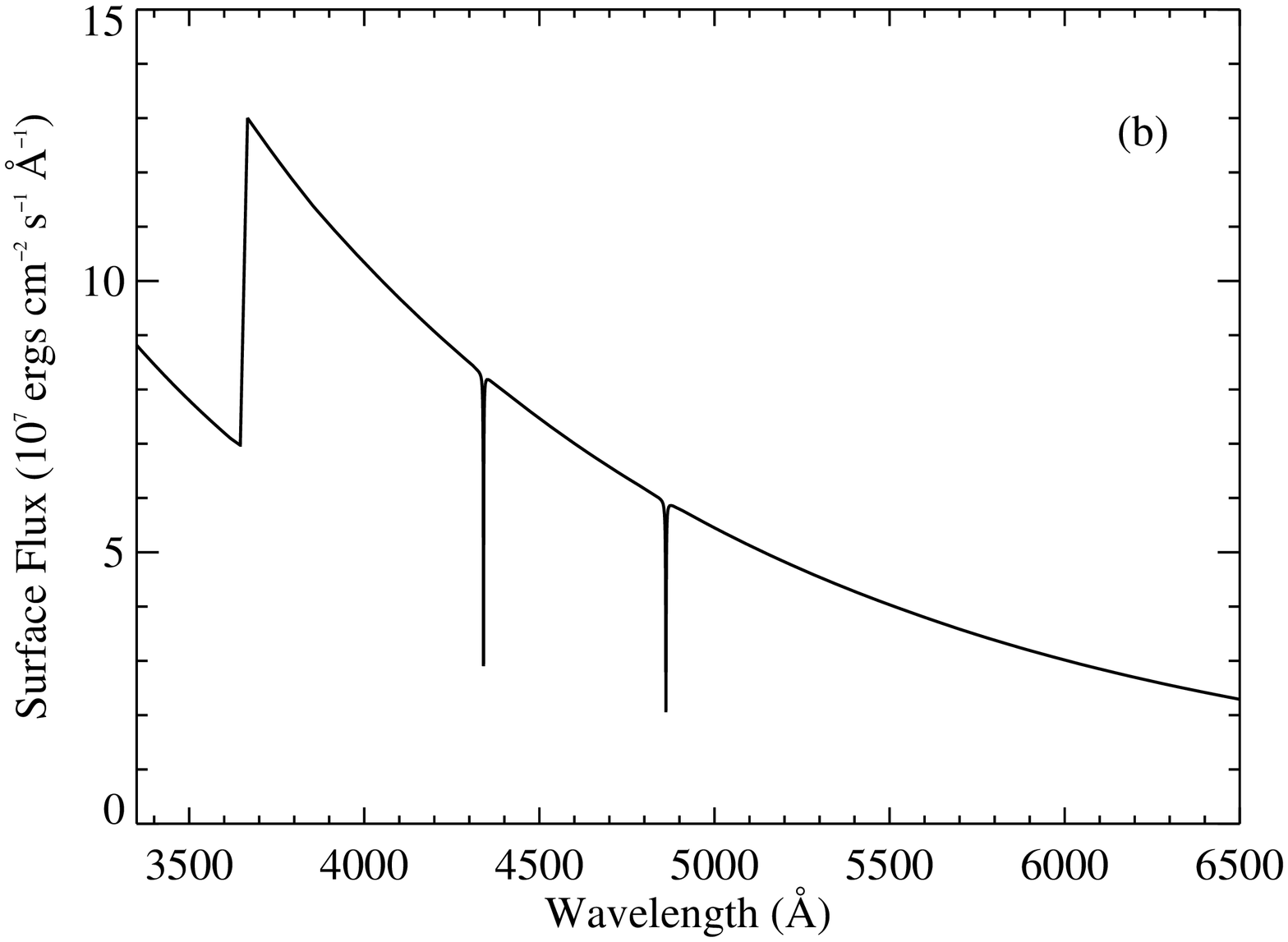} 

 \caption{(a) The quiescent M dwarf atmosphere from \cite[Allred et al. (2006)]{Allred2006} with a phenomonological hot spot below
the temperature minimum region. (b) The static code RH is used to calculate the emergent flux spectrum for a 6-level Hydrogen atom. The slope of this continuum closely resembles the continuum of the subtracted flare spectrum.  Moreover, the temperature bump generates absorption features from Hydrogen. }
   \label{fig2}

\end{figure}


\begin{discussion}

\discuss{Kosovichev}{1) Have you been able to observe Doppler shifts?  2) Some theoretical models
suggested that condensations behind the downward propagating shock may be responsible for white
light emission.  What is the satus of these models?  Can this be ruled out by the new observations?}

\discuss{Kowalski}{First, our spectra have a low spectral resolution, with R$\sim$1000.  We aren't concerned about getting very accurate wavelength calibration because we don't want to go off the slit very often to get
an arc exposure, in case there is a flare.  Flare rise times are very fast, typically 20-40s, so we don't want to miss it.  We are more concerned with getting the flare and accurate flux calibration than with the wavlength calibration.

These RADYN models predict a downward condensation wave with speeds of tens of km per second, but this
doesn't reach the photosphere with sufficient energy.  Most of the energy that reaches the photosphere is
backwarming radiation from the Balmer continuum.  I'd just like to add that X-ray backwarming was also once 
thought to be a candidate for the heating in deep layers;  however, according to these models only $\sim$1\% of the heating is caused by
X-ray backwarming.  }

\end{discussion}

\end{document}